# Phylogeny.fr: the phylogenetic platform designed for non-specialists

Valentin Gorgodian[1], Olivier Poirot[1], Alain Schmitt[1], Virginie Collomb[1], Jean-Michel Claverie[1], Chantal Abergel[1], Matthieu Legendre[1], Sebastien Santini[1#]

**Affiliation**

[1]Information Génomique & Structurale, UMR7256, IMM, IM2B, IO, CNRS & Aix-Marseille Université, Marseille, France

[#]Corresponding author: sebastien.santini@igs.cnrs-mrs.fr

**Running head**

Phylogeny.fr 2.0: major updates and new features

**Abstract**

Phylogenetic analysis has become a standard approach across many areas of biology, yet the growing complexity of phylogenetic methods and software remains a major obstacle for non-specialists. Since its launch in 2008, Phylogeny.fr has provided an accessible web platform for building phylogenetic trees using widely accepted methods without requiring local software installation. Here, we present a major redesign and modernization of the service. The new version integrates state-of-the-art tools while preserving historical programs for legacy support and relies on modern web architecture and HPC infrastructure. New interactive React-based viewers, ReSeqt and Reactree, provide intuitive exploration and publication-ready visualization of alignments and trees. The Blast-Explorer companion tool has also been updated and now includes clustering options. By combining ease of use, methodological flexibility, and modern phylogenetic tools, the new Phylogeny.fr addresses the needs of researchers, teachers, and students seeking accessible and reliable phylogenetic analyses.

## 1 Introduction

Constructing a phylogenetic tree from molecular biology data (primarily gene or protein sequences) has become a standard practice in the completion of a growing number of biological research projects, far beyond their traditional use by evolutionary biologists. At the same time, the theoretical and statistical foundations of the algorithms used have become vastly more complex, ranging from the simple, intuitive nearest-neighbor method [1] to Bayesian [2] and maximum-likelihood approaches [3].

Building a phylogenetic tree requires four distinct steps: - 1) identify and download a set of homologous DNA or protein sequences, - 2) generate an optimal multiple alignment of those sequences, - 3) build a tree from the aligned sequences, and -4) generate a publication-ready graphical representation of that tree clearly conveying the relevant information. Identifying the software needed for these various steps, selecting the most widely accepted ones, and installing them locally is often beyond the capabilities of many laboratories that only use phylogenetic trees on a very occasional basis. The goal of the Phylogeny.fr platform, launched in 2008 [4] , was to enable non-specialists to choose among this suite of applications and use it without having to install anything on their own computers, while ensuring that they were using the most widely accepted methods. It has since become one of the most cited online phylogenetic platforms.

Although designed for the non-specialist in mind, it is yet flexible enough to address expert questions. For this, the original three workflows have been conserved and allow users to perform analysis ranging from fully automated pipelines to customizable expert analysis. For instance, four different multiple sequence alignment tools are available (MAFFT [5], Muscle [6], clustal omega [7] and T-Coffee [8]) to optimally handle various data sets. In 2010, blast-explorer [9] was developed and integrated in phylogeny.fr as a companion tool to build datasets of homologous sequences suitable for phylogenetic analysis. In 2019, a fork of the historical version of the service was released [10] including an update of the phylogeny software but leaving aside some competitors in the field [3, 11].

When the first version of phylogeny.fr was created, a large number of models, methods, and programs already existed for phylogenetic tree inference [4]. Since then, new software has emerged that combines some of these methods and offers even more models. At the same time, genome sequencing technologies have become even more powerful, and the number of available genomes and sequences has exploded. For example, the number of entries in RefSeq has risen from 5.6 million in 2008 [12] to over 450 million in 2024 [13]. Consequently, the number of studies requiring at least a small phylogenetic analysis has increased accordingly as well as the number of biologists lost in this complex environment. Phylogeny.fr was developed to meet their needs without requiring them to master all the parameters necessary for the calculations. In addition to its use in research, phylogeny.fr user-friendly interface has made it a tool of choice for teaching, and it is now an essential component of many educational programs. But this interface is now outdated in terms of both design and functionality and was in dire need of an overhaul.

Here we present a major upgrade of our historical service integrating state-of-the-art alignment and tree inference tools together with a redesigned and modernized web interface. The most cutting-edge programs have replaced the outdated ones that were previously in use. The latter have been preserved for legacy. The client side has been optimized for speed and the job distribution to fit the specification of a dedicated cluster.

Phylogeny.fr remains one of the most widely used and cited phylogeny services, catering to non-specialists, experts, and students alike. This update is part of our commitment to quality and provides access to both the most advanced software in the field and historical programs. The user experience has been enhanced through the use of modern web technologies. While maintaining the same spirit as during the original development, we hope to integrate structural information in the near future to offer more comprehensive phylogenies.

## 2 Material

The service was completely redesigned to fit the HPC infrastructure and modern web standards. The Phylogeny.fr web interface is now running on a dedicated virtual machine all computation jobs submitted to a dedicated computing cluster using slurm as batch scheduler and environment modules to manage program versions.

The frontend is built with Next.js 15 (React 19, TypeScript), which also serves as the backend API layer. Several steps are handled client-side including input validation, sequence-type detection, and result visualization, to reduce server load.

The service is protected against abuse through a Redis-backed hybrid rate limiter operating at both session and IP levels, and all tool commands are regenerated server-side from validated parameters to prevent injection attacks.

Input limitations depend on the selected program and its computational efficiency, as shown in Table 1. All alignment tools (MAFFT, Muscle, Clustal Omega, T-Coffee) and trimming tools (Gblocks, ClipKit, BMGE) accept up to 200 sequences. Among phylogeny programs, IQ-TREE, RAxML-NG and FastTree are similarly limited to 200 sequences, while PhyML, MrBayes and BioNJ are restricted to 50 sequences given their higher computational cost. Sequence length is capped at 6,000 bases for nucleotide sequences and 2,000 residues for protein sequences across all tools.

## 3 Methods

All modern programs that work with biological sequences (amino acids and nucleotides) are compatible with the FASTA format, which is by far the most widely used format. Therefore, we have decided to allow only this format as input for programs that manipulate sequences (whether aligned or not). However, to ensure broad compatibility, this new version of phylogeny.fr also offers a conversion tool from and to several other formats. Similarly, only the Newick format is supported for phylogenetic trees.

The three pipelines that made phylogeny.fr famous have retained their names but have been redesigned to better align with the usage patterns observed during the service's years of operation. State-of-the-art programs are now available; the versions installed at the time of publication are not final, and they will be updated as frequently as possible. The list of settings selected for each program that the user can adjust is detailed in the service documentation section.

### 3.1 “One Click” mode

The “One Click” mode was originally designed to be as simple as possible: just copy your raw sequences and run the calculations. While retaining the spirit of this mode, we now offer the option to choose between faster or more accurate computation, requiring only one additional click. The input data are an unaligned multiple sequences fasta file. The default multiple alignment program is now MAFFT [5] with its default parameters, chosen for its speed and its ability to handle large samples. By default, the poorly aligned regions presenting too many gaps are removed with ClipKIT [14] (with options --mode gappy and --gaps 0.4). This step can be disabled. If the “fast” execution mode is selected, the tree is inferred using FastTree [15] with its default parameters which uses heuristics to speed up the computation. The “accurate” execution mode uses an highly cited and commonly used ML phylogenetics software, IQ-Tree [16] (with -bb 5000 -bi 200). The tree obtained is automatically midpoint rooted and displayed. It can be downloaded as a newick file.

### 3.2 Advanced mode

Advanced mode uses exactly the same programs as “One Click” Accurate but allows to edit their settings. Only a selection of the most relevant settings for each program is available to keep the interface simple and the workflow easy to use.

### 3.3 “A la carte” mode

This mode allows to create a custom workflow using various custom programs and edit their relevant settings. The sequences alignment is a mandatory step that can be performed with MAFFT [5] or one of the following programs: Muscle [6] , clustal omega [7] or T-Coffee [8]. The alignment curation step is optional but recommended and can be performed using BMGE [17], Gblock [18] or the more recent ClipKit [14]. The tree inference step is also mandatory and can be performed with PhyML [19], BioNJ [20], Mrbayes [21] or recently developed programs such as FastTree [15], RAxML-ng [22] or IQ-Tree [16].

### 3.4 Alignment and tree viewers

Special care has been taken with the tools for visualizing trees and alignments and two independent React modules have been developed from scratch. More than just visualization tools, they allow to explore trees and alignments in an intuitive and seamless way, manipulating them to produce clear, clean, and publish-ready results. Moreover, they have been coded as light and fast standalone components and can be integrated in any React environment with one import.

ReSeqt (Fig. 1) is the new alignment viewer. It can display sequences in different orders: as generated by the alignment program, classified by name, length, percentage of gap, identity compared to a selected sequence in the alignment or ordered by hand. The alignment can be colored by residue property, conservation of a site or identity compared to a selected sequence. The conservation rate and the consensus, defined by the most represented residue at each site can also be displayed. The entire alignment can be browsed using the overview and a dedicated scroll bar. A user selected part of the alignment can be exported as fasta file or png file. The repository is available at https://github.com/vaaloo/reseqt

Reactree (Fig. 2), the tree viewer, accepts a newick formated file as main input. The tree can be displayed rectangular or circular, as a phylogram or a cladogram with the leaf tags aligned or not. A toolbar allows to re-root the tree on a selected node or at midpoint as well as flip, swap or collapse branches/nodes. Branch color and thickness can be set and the tag of collapsed branched can be edited. Bootstrap or branch length can be displayed. From both a teaching and research perspective, Reactree allows users to display the multiple alignment used to generate the tree following the sequence identifiers. This makes it possible, for example, to quickly identify which type of substitution or indel is responsible for the tree's topology without the need to open the alignment in another window. A search button allows you to quickly find a given taxon within a complex tree. The whole tree can be exported as a picture (jpeg, png, svg) but also as a newick format retaining the tree conformation, after a re-rooting for example. The repository is available at https://github.com/vaaloo/reactree.

### 3.5 Blast Explorer

This companion tool of phylogeny.fr has also been modernized. The four standard BLAST procedures (BLASTp, BLASTn, tBLASTn, and BLASTx) have been retained and allow users to search multiple databases. These databases were selected based on their size, which is manageable on our computing infrastructure, and may be subject to change depending on hardware upgrades to the service. The preliminary list includes: clusteredNR that is a clustered version (90% identity / 90% length) of the NCBI protein NR database, swissprot, PDB (amino acids and nucleotides sequences), Uniref90, RefSeq Viral (protein and genomic sequences), Core nucleotide database. The E-value threshold can be set from 1 to $1e^{-100}$.

The number of homologous sequences the service can provide is intentionally limited to 500, of which 100 will be displayed in a tree. Depending on the database selected, the returned homologues may be too closely related to be informative. To address this potential issue, we have added the option to cluster these sequences using MMseqs2 [23]. It is possible to adjust the percentage of identity and the target/query ratio coverage to control the process. The first 100 sequences returned by blast, clustered or not, are then aligned using mafft and a tree is calculated using fastTree with their respective default options. This interactive tree (Fig. 3A) can be used to select a subset of sequences by clicking each taxa or based on criteria presented in the “Tree View” toolbox (Fig. 3B). The “All hits” toolbox (Fig. 3C) offers the options to select sequences using more criteria among all results below the desired e-value threshold. The selected sequences can be sent to “One Click” or advanced workflows.

### 3.6 References and history:

Phylogeny.fr relies on numerous programs to provide the requested results. Therefore, we believe it is important that, whenever possible, in addition to the service itself, all of the programs used should be cited. To make this task easier for users, after each run (workflow or standalone tool), a ready to publish small text is generated including all tool references used for the job. Only edited options must be added by the user.

The first version of the site allowed users to create an account and save their results. Over the many years the service has been in operation, it became clear that this option was underutilized, despite increasing administrative complexity, particularly due to restrictions surrounding personal data. It was therefore decided to remove this option from the site. Since server storage capacity is limited, data is kept accessible for only one week and archived for one month. However, the technology used for coding has made it possible to establish a job history for each user, which is stored in their browser cache and retained for one week. Furthermore, when a user submits its email address to receive results, it is retained only for the duration of the job and deleted as soon as the job completion confirmation email has been sent. No sensitive data is therefore stored on the server.

**Aknowledgements**

We would like to extend our special thanks to all the authors who develop and maintain the tools used in Phylogeny.fr. The PACA Bioinfo platform maintains the website and the servers that handle all the calculations. Valentin Gorgodian was supported by IA ReNaBi-IFB (ANR-11-INBS-0013).

## Figures

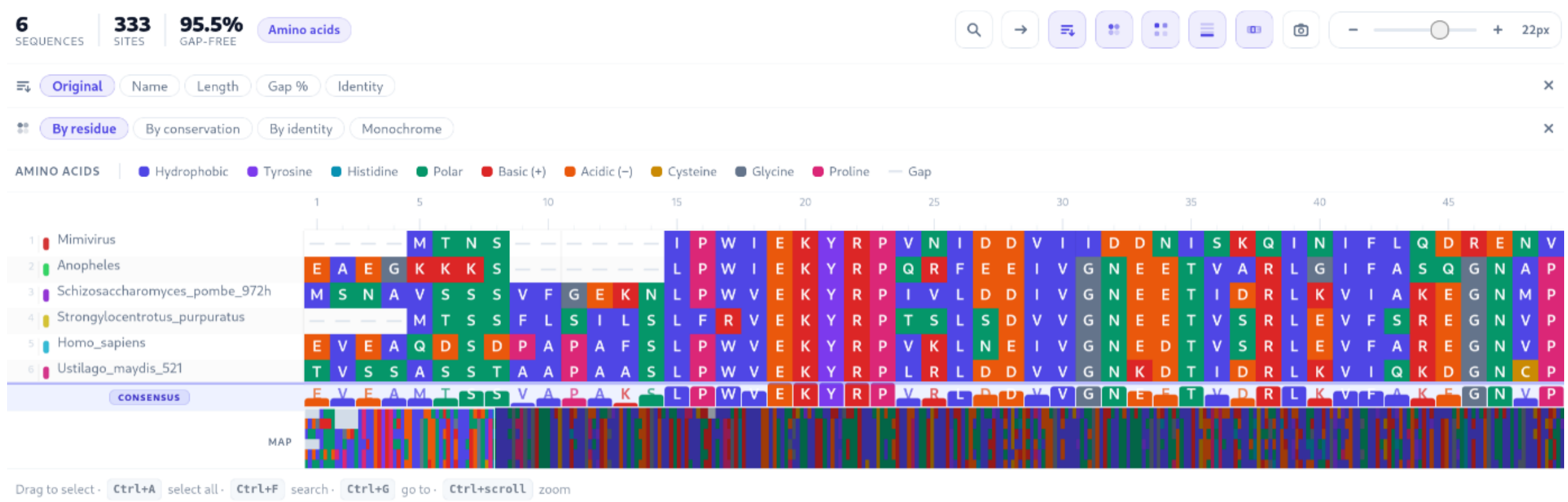

Fig. 1: ReSeqt, the alignment viewer consists of several tracks. The left side of the first track displays information about the alignment, and the right side is dedicated to features. From left to right, there are 8 buttons to search for a sequence by its identifier, jump to a specific position in the alignment, sort the alignment by various criteria, change the color scheme, show/hide the color legend track, show/hide the consensus track below the alignment, show/hide the overview below the alignment, export as PNG, and a slider to change the font size. One more button permits to show/hide a conservation track above nucleotide alignments.

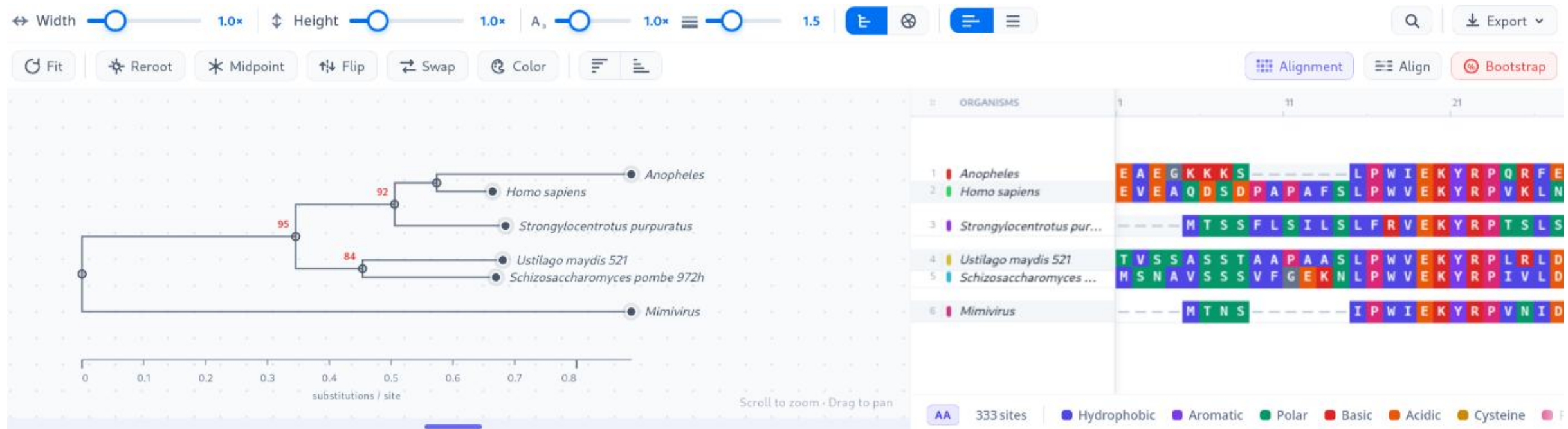

Fig. 2: Reactree, the new tree viewer. The first track contains sliders for adjusting the tree's width and height, the font size, and the thickness of the branches. The first track contains sliders for adjusting the tree's width and height, the font size, and the thickness of the branches, as well as buttons for switching between rectangular and circular representations or between a dendrogram and a cladogram, the search button and the export button. The second track contains all commands acting on the tree configuration: re-root on a selected node, midpoint root, flip or swap branches or reorder branches by clade size. It also includes buttons for changing the colors of the branches, managing the presence of the corresponding multiple alignment, aligning the identifiers, and displaying the bootstrap or branch length values.

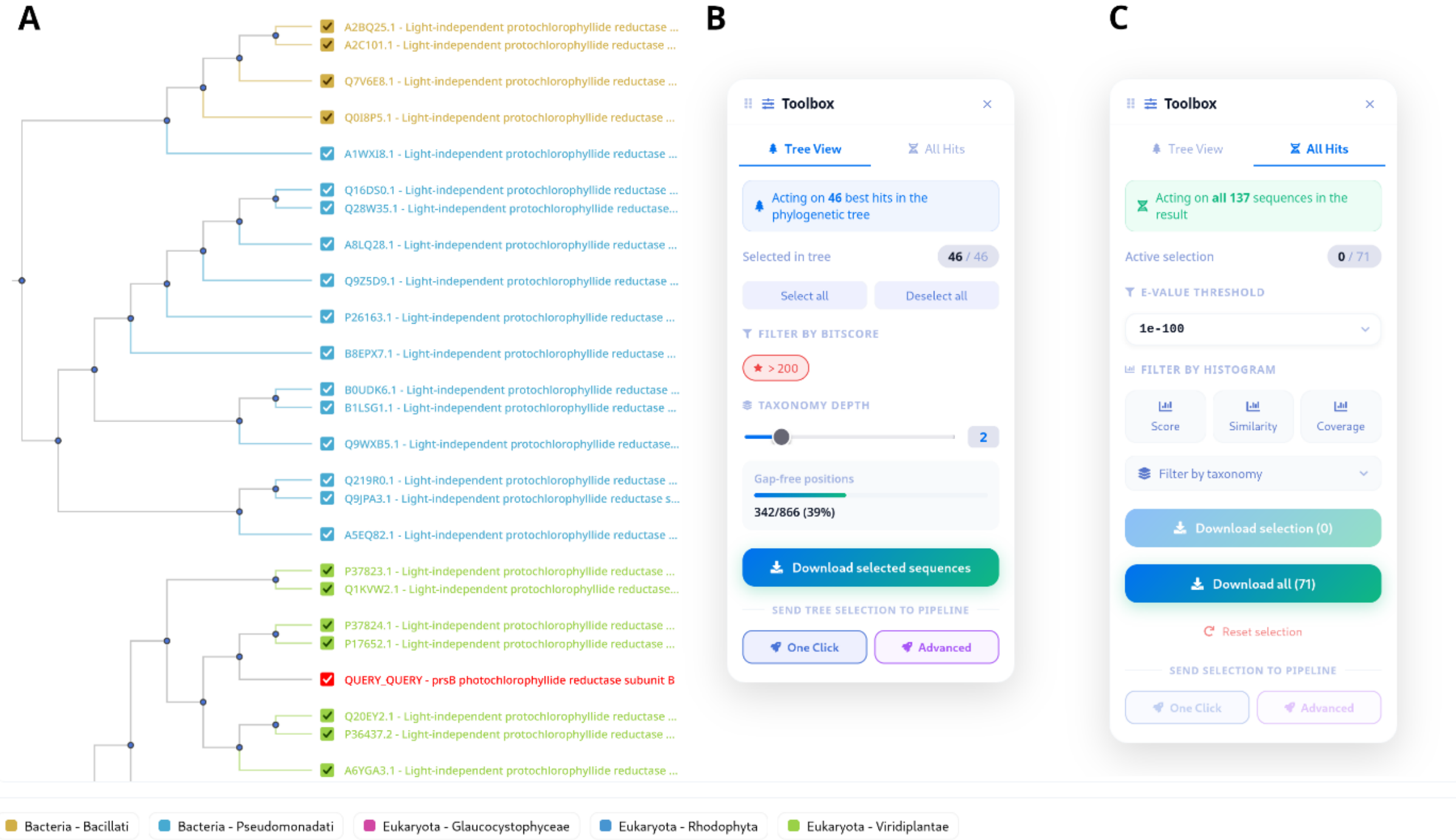

Fig. 3: Example of blast explorer output. A) In the rendered tree, the taxa are colored depending on the taxonomy depth and can be selected/deselected by simple click. B) The "Tree View" toolbox only allows user to select/deselect all taxa with one button, filter the results by bit score or change the taxonomy depth from 1 (domain) to 8. C) In the "All-Hits" toolbox, the set of sequences from which selections can be made is defined by adjusting the e-value threshold. Sequences can then be selected using histograms based on score, similarity, or coverage parameters. It is also possible to filter according to a specific taxonomy.

| Tool | Description | Version | Max Sequences |
| --- | --- | --- | --- |
| Blast | Basic Local Alignment Search Tool | 2.16.0 | 1 |
| MMseqs2 | Ultra-fast and sensitive sequence search and clustering | sse2 | — |
| Mafft | Multiple sequence alignment program for large datasets | 7.526 | 200 |
| Muscle | Fast and accurate multiple sequence alignment | 5.3 | 200 |
| Clustal Omega | Fast multiple sequence alignment | 1.2.4 | 200 |
| TCoffee | Advanced multiple sequence alignment program | 13.46.0.919e8c6b | 200 |
| GBlocks | Alignment curation — eliminates poorly aligned and divergent regions | 0.91b | 200 |
| ClipKit | Alignment curation using smart-gap trimming | 2.11.4 | 200 |
| BMGE | Alignment curation using gap trimming | 2.0 | 200 |
| FastTree | Approximately-maximum-likelihood phylogenetic tree inference | 2.2.0 | 200 |
| IQ-Tree | Maximum-likelihood phylogenetic inference with automatic model selection | 2.4.0 | 200 |
| RAxML-NG | Randomized Axelerated Maximum Likelihood phylogenetic inference | 2.0.0 | 200 |
| MrBayes | Bayesian inference of phylogenetic trees | 3.2.7a | 50 |
| PhyML | Maximum-likelihood phylogenetic inference | 3.3.20241207 | 50 |
| Fastphylo | Fast tools for phylogenetics — distance computation and neighbor-joining | 1.0.1 | — |
| BioNJ | Neighbor-joining phylogenetic inference | 1.0 | 50 |

Table 1: List of available programs, including their versions, functions, and limits on the number of sequences